\documentclass[sigconf]{acmart}
\pdfoutput=1 
\usepackage{amsmath,amsfonts,amssymb}
\usepackage{eqnarray}
\usepackage{mathtools}
\usepackage{balance} 
\usepackage{mathrsfs}
\usepackage{multirow}
\usepackage{csquotes}
\usepackage{bbm}
\usepackage{enumitem} 

\usepackage[]{algorithm2e}
\usepackage{graphics}
\usepackage{subcaption} 
\usepackage{units}
\usepackage{csquotes}
\usepackage{booktabs}
\usepackage{lscape}
\usepackage{changepage}

\copyrightyear{2018}
\acmYear{2018}
\setcopyright{acmcopyright}
\acmConference[WSDM 2018]{WSDM 2018: The Eleventh ACM International Conference on Web Search and Data Mining}{February 5--9, 2018}{Marina Del Rey, CA, USA}
\acmBooktitle{WSDM 2018: The Eleventh ACM International Conference on Web Search and Data Mining, February 5--9, 2018, Marina Del Rey, CA, USA}
\acmPrice{15.00}
\acmDOI{10.1145/3159652.3159689}
\acmISBN{978-1-4503-5581-0/18/02}
\newcommand{\paragraphHd}[1] {\vspace{1.2mm}\noindent\textbf{#1.}} 
\newcommand{\ra}{\renewcommand{\arraystretch}{1.2}}

\makeatletter
\def\@fnsymbol#1{\ensuremath{\ifcase#1\or *\or \dagger\or \ddagger\or
   \mathsection\or \mathparagraph\or \|\or **\or \dagger\dagger
   \or \ddagger\ddagger \else\@ctrerr\fi}}
\makeatother

\title[Co-PACRR]
{Co-PACRR:\\A Context-Aware Neural IR Model for Ad-hoc Retrieval}

 \author{Kai Hui}
 \affiliation{%
   \institution{Max Planck Institute for Informatics /}
   \institution{SAP SE}
 }
 \email{kai.hui@sap.com}
 \author{Andrew Yates}
 \affiliation{%
   \institution{Max Planck Institute for Informatics}
 }
 \email{ayates@mpi-inf.mpg.de}
 \author{Klaus Berberich}
 \affiliation{%
   \institution{Max Planck Institute for Informatics /}
   \institution{htw saar}
 }
 \email{kberberi@mpi-inf.mpg.de}
 \author{Gerard de Melo}
 \affiliation{%
   \institution{Rutgers University--New Brunswick}
 }
 \email{gdm@demelo.org}

\begin{abstract}
Neural IR models, such as DRMM and PACRR,
have achieved strong results by successfully
capturing relevance matching signals.
We argue that the context of these matching signals
is also important.
Intuitively, when extracting, modeling, and combining matching signals,
one would like to consider the surrounding text
(local context) as well as other signals from the same document that can
contribute to the overall relevance score.
In this work, we highlight three potential shortcomings caused by not considering context information
and propose three neural ingredients to address them:
a disambiguation component, cascade k-max pooling,
and a shuffling combination layer.
Incorporating these components into the PACRR model yields Co-PACRR,
a novel context-aware neural IR model.
Extensive comparisons with established models
  on \textsc{Trec} Web Track data
  confirm that the proposed model can achieve superior search results.
  In addition, an ablation analysis is conducted to gain insights
  into the impact of and interactions between different components.
We release our code to enable future comparisons\footnote{\url{https://github.com/khui/repacrr}}.
\end{abstract}

\begin{document}
\maketitle

\section{Introduction} 
\label{sec.introduction}
 \begin{figure*}
 \centering   
 \includegraphics[width=\linewidth]{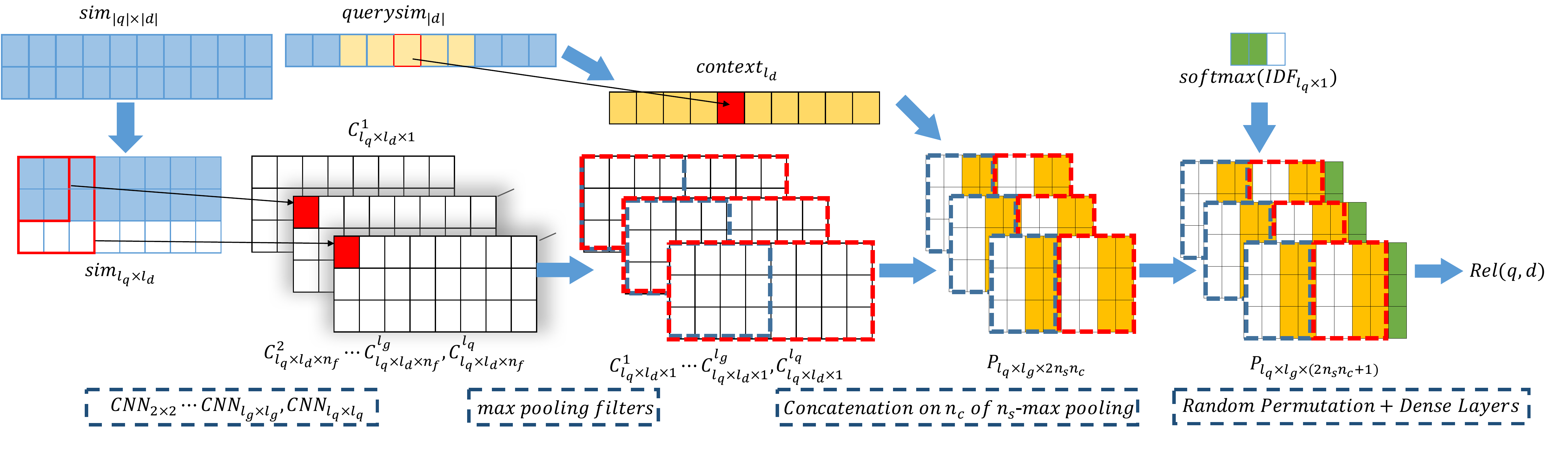}
 \caption{The pipeline of Co-PACRR. 
 The inputs include two matrices, namely, $\mathit{sim}_{|q|\times |d|}$ and 
 $\mathit{querysim}_{|d|}$.
 All these similarity matrices are truncated/zero-padded to the dimensionalities governed by 
 $l_q$ and $l_d$. 
Several 2D convolutional kernels are first applied to the similarity matrices,
one for each $l_g\in[2,l_g]$.
Next, max pooling is applied to the filters, leading to $l_g+1$ matrices, namely,
$C^1\cdots C^{l_g}, C^{l_q}$.
   Following this,
   $n_s$-max pooling captures the strongest $n_s$ signals on each $C$, 
   at all $n_c$ positions from $cpos$.
   At the same time, the context similarity corresponding to each term in $top$-$n_s$ from $\mathit{context}_{l_d}$
is also appended, leading to $P_{l_q\times l_g \times (2 n_s n_c)}$.
   Finally, the query terms'
   normalized IDFs are appended, and a feed forward network is applied,
   after permuting the rows in $P_{l_q\times l_g \times (2 n_s n_c+1)}$,
   yielding a query-document relevance score $\mathit{rel}(q,d)$.
In this plot, a setting with $l_d=8$, $l_q=3$, $l_g=3$,
$n_c=2$, $n_s=2$, and $cpos=[50\%,100\%]$ is shown. 
 }\label{fig.model}
 \end{figure*}

State-of-the-art neural models for ad-hoc information retrieval 
aim to model the interactions between a query and a document to produce a relevance score,
which are analogous to traditional interaction signals such as BM25 scores.
Guo et al.~\cite{guo2016deep} 
pointed out that a neural IR model 
should 
capture query-document interactions in terms of
\textit{relevance matching} signals 
rather than capturing \textit{semantic matching} signals
as commonly used in natural language processing (NLP).
Relevance matching focuses on the 
pertinence of local parts of the document with respect to the query (e.g., via n-gram matches),
whereas semantic matching captures the overall semantic similarity between the query and the entire document.
Accordingly,
relevance matching over unigrams
has been successfully modeled using histograms in the \textit{DRMM} model~\cite{guo2016deep},
using a convolutional layer in \textit{DUET}'s local model~\cite{mitra2017learning},
and using a pool of kernels in the more recent \textit{K-NRM} model~\cite{xiong2017end}.
In addition,
position-aware relevance matching signals
are further captured in \textit{PACRR}~\cite{hui2017position} with the goal of encoding matching signals beyond unigrams,
such as n-gram matches and ``soft'' n-gram matches, in which the order of some terms is modified.

Existing models have achieved strong results by focusing on modeling relevance matching signals.
However, we argue that the context of such signals are also important but has yet to be fully accounted for in these models.
Intuitively, a matching signal contributes to the final relevance score 
within the context of 
its local text window and the context of all matching signals from the whole document.
Given a matching signal, 
a text window that embeds the signal is referred to as its \textit{local context}, whereas
all matching signals from the same document
are referred to as the signal's \textit{global context}.
Inspired by past research within the IR community,
we first highlight three particular 
shortcomings that can be addressed by incorporating context. Thereafter,
we introduce novel neural components to address the shortcomings within
PACRR~\cite{hui2017position}, a state-of-the-art neural IR model.
This ultimately leads to Co-PACRR (context-aware PACRR), a novel model as summarized in Figure~\ref{fig.model}.

To start with, when disregarding the local context,
the matching signals extracted
between terms from a query and a document
may suffer from \textit{ambiguity}.
For example, in the query
``\textit{Jaguar SUV price}'', the term ``Jaguar'' refers to a car brand, but
``Jaguar'' also happens to be the name of a species of animal.
Such ambiguity 
can mislead a model to extract false positive matching signals.
In the above example, an occurrence of the term ``jaguar'' referring to the animal
should not contribute much to the document's relevance score.

Beyond this, accounting for the global document context may be important as well.
Some such signals are desirable, while others need to be disregarded.
In particular, we conjecture that the \textit{the location of the matches}
is important to better account for
the level of reading effort needed to reach 
the relevant information.
For example,
consider two pseudo-documents that are both
concatenations of one relevant and one
non-relevant document, but in a different order.
Although the same relevant information is present,
extra effort is required when reading the pseudo-document
where the non-relevant document appears first.

Not all aspects of the document context, however, are beneficial. In particular, 
we argue that the order in which the document matches different query terms may vary, as there can be many ways to express the same information request.
When combining matching signals from
different query terms, PACRR employs a recurrent 
layer,
whereas DRMM, K-NRM, the local model in DUET,
and MatchPyramid employ feedforward
layers.
Both kinds of models may be sensitive to the order in which query terms are matched,
as the signals from individual query term matches and 
their associated positions in a query are jointly considered.
Learning a query term order-dependent combination is particularly concerning when
the query dimension is zero-padded
(as in most models),
because the aggregation layer may incorrectly learn
to down-weight the positions that are zero-padded more often (e.g., at the end of 
a short query) in the training data.
More generally, the aggregation layer may learn to treat matching signals differently depending on the position of a term within the query. This may hurt the model's ability to generalize to different reformulations of a given query, and it is also unnecessary because positional information is already accounted for in an earlier layer.

To close these gaps, 
we introduce neural components to
cater to both the local and the global context.
Intuitively,
to avoid extracting false positive matching signals due to ambiguity,
matching signals are double-checked based on their local context and
penalized if there is a mismatch between the senses of words between the document and the query.
To consider the global context of matching signals, the signals' strengths at different
document positions are considered.
To disregard the absolute positions of terms in the query,
the sequential dependency over query terms 
is decoupled before the aggregating combination layer.
While these ideas apply more generally,
we incorporate them into the PACRR architecture to develop specific neural components,
which leads to the Co-PACRR model that contains the following new components:
\begin{itemize}
 \item[-] A \textit{disambiguation} building block
to address the challenge of ambiguity
by co-considering salient matching signals 
together with the local text window in which they occurred.
\item[-]
A \textit{cascade k-max pooling} approach
in place of regular k-max pooling layers,
enabling the model 
to benefit from information about the location of
matches.
These locations 
are jointly modeled together with the matching signals.
This is inspired by the cascade model~\cite{craswell2008experimental},
which is based on the idea that relevance gains
are influenced by the amount of relevant 
information that has already been observed. 
\item[-]
A \textit{shuffling combination} layer
to regularize the model so as to disregard the absolute positions of terms
within the query.
Removing query-dependent context before combination
improves the generalization ability of 
the model.
\end{itemize}

\paragraphHd{Contributions}
We incorporate the aforementioned building blocks
into the established PACRR model,
leading to the novel Co-PACRR model,
jointly modeling matching signals with their local and global context.
Through a comparison
with multiple state-of-the-art models including 
DRMM, K-NRM, the local model in DUET,
MatchPyramid, and the PACRR model
on six years of
\textsc{Trec} Web Track benchmarks,
we demonstrate 
the superior performance of Co-PACRR.
Remarkably, 
when re-ranking the search results from a na\"ive initial
ranker, namely a query-likelihood ranking model,
the re-ranked runs are ranked within the top-$3$ on at least five years 
based on ERR@20.
In addition,
we also investigate 
the individual and joint effects of the proposed components
to better understand the proposed model in an \textbf{ablation analysis}.

\paragraphHd{Organization} The rest of this paper unfolds as
follows. 
We discuss related work in Section~\ref{sec.relatedwork} and put our work in context.
Section~\ref{sec.background} recaps the basic neural-IR model PACRR,
and thereafter 
Section~\ref{sec.method} describes the proposed building components
in detail.
The setup, results, and analyses of our extensive experimental evaluation
can be found in Section~\ref{sec.evaluation} and Section~\ref{sec.discussion}, before
concluding in Section~\ref{sec.conclusion}.


\section{Related Work} 
\label{sec.relatedwork}
In ad-hoc retrieval, a system aims at 
creating a ranking of documents according to their
relevance relative to a given query. 
The recent promises of deep learning methods
as potential drivers for further advances in retrieval quality have attracted significant attention.
Unlike learning-to-rank methods, where models are learned on top of 
a list of handcrafted features~\cite{liu2009learning},
a neural IR model aims at modeling the interactions between a query and a document
directly based on their free text.
Actually, the interactions being learned in a neural IR model 
correspond to one of the feature groups employed in learning-to-rank methods.
They involve both a query and a document, as do
BM25 scores. 
The proposed Co-PACRR belongs to this class of neural IR models
and is hence compared with other neural IR models in Section~\ref{sec.evaluation}.

As described in Section~\ref{sec.introduction},
neural IR approaches can be categorized as \textit{semantic matching} and \textit{relevance matching} models. 
The former follows the embedding approach adopted in many natural language processing tasks,
aiming at comparing the semantic meaning of two pieces of text
by mapping both into a low-dimensional representation space.
Therefore, models developed for natural language processing tasks
can also be used as retrieval models by assigning a similarity score 
to individual query-document pairs.
For example, 
ARC-I and ARC-II~\cite{Pang:2016:TMI:3016100.3016292}
are two such models developed for
the tasks of sentence completion, identifying the response to a microblog post, and performing
paraphrase detection. 
In addition,
Huang et al.~\cite{Huang:2013:LDS:2505515.2505665} proposed Deep Structured Semantic Models (DSSM),
which learn low-dimensional representations of a query and a document in a semantic space
before evaluating the document
according to its cosine similarity relative to the query.
Similar approaches such as 
C-DSSM~\cite{Shen:2014:LSR:2567948.2577348} further
employed alternative means to learn dense representations of the documents.

In comparison,
Guo et al.~\cite{guo2016deep} argued that the matching
required in information retrieval is different from the matching used in NLP tasks,
and that \textit{relevance matching} is better suited for retrieval tasks.
Relevance matching
compares two text sequences jointly, namely, a document and a query,
by directly modeling their interactions.
In relevance matching, local signals such as unigram matches are important.
Meanwhile, semantic matching seeks to model the semantic meaning
of the two text sequences independently, 
and the matching is considered in a semantic space.
Accordingly, the
Deep Relevance Matching Model (DRMM)~\cite{guo2016deep} 
was proposed to model unigram relevance matching by
encoding a query-document pair in terms of a histogram of similarities between terms
from the query and the document.
More recently,
K-NRM~\cite{xiong2017end} relied on a pool of kernels in place of the 
histogram, capturing the unigram relevance matching in a more smooth manner, addressing
the issues of bin boundaries in generating histograms.
In addition to the unigram signals,
position-aware neural IR models have been proposed, such as 
MatchPyramid~\cite{NIPS2014_5550,DBLP:journals/corr/PangLGXC16},
which is motivated by works from computer vision~\cite{simonyan2014very},
 and PACRR~\cite{hui2017position},
 which follows the ideas of term dependency~\cite{DBLP:conf/cikm/HustonC14,metzler2005markov} and query term proximity~\cite{tao2007exploration} modeling 
 from ad-hoc retrieval.
 Both encode matching signals beyond a single term with
  convolutional neural networks (CNNs).
 Beyond that,
Mitra et al.~\cite{mitra2017learning} proposed DUET, 
a hybrid deep ranking model combining
both kinds of matching,  
with two independent building blocks, namely,
a local model for relevance matching and a distributed model
for semantic matching. 
The proposed Co-PACRR model
belongs to the class of relevance matching models, and
attempts to further incorporate the context of matching signals.

\section{Background} 
\label{sec.background}

In this section, we summarize the PACRR model~\cite{hui2017position},
which we build upon by proposing novel components.
When describing PACRR, we follow the notation from~\cite{hui2017position}.
In general, PACRR takes a similarity matrix between a query $q$ and 
a document $d$ as input, and the output of the model is a scalar,
namely, $\mathit{rel}(d, q)$,
indicating the relevance of document $d$ to query $q$.
PACRR attempts to model query-document interactions based on these similarity matrices.
At training time, the relevance scores for 
one relevant and one non-relevant document,
denoted as $d^+$ and $d^-$, respectively, are fed into a
max-margin loss as in Eq.~\ref{eq.maxmarginloss}.
\begin{equation}\label{eq.maxmarginloss}
 \mathcal{L}(q,d^+, d^-;\Theta)=\mathit{max}(0,1-\mathit{rel}(q,d^+)+\mathit{rel}(q,d^-))
\end{equation}
In the following, PACRR is introduced component-by-component.
\begin{enumerate}
\item \textbf{Input}: the similarity matrix $\mathit{sim}_{l_q\times l_d}$,
where both $l_q$ and $l_d$ are hyper-parameters unifying the 
dimensions of the input similarity matrices. 
$l_q$ is set to the longest query length,
and $l_d$ is tuned on the validation dataset.
Given the settings for both $l_q$ and $l_d$,
a similarity matrix between a query and a document is truncated 
or zero-padded accordingly;
 \item \textbf{CNN kernels 
 and max-pooling layers}: multiple CNN kernels with $l_f$ filters 
 capture the query-document interactions, like n-gram matching, 
 corresponding to different text window lengths, namely $2, 3, \cdots, l_g$.
 The hyper-parameters $l_g$ and $l_f$ govern the longest text window under consideration 
 and the number of filters, respectively.
 These CNN kernels are followed by a 
 max-pooling layer on the filter dimension to retain the strongest matching signal for each kernel,
leading to $l_g$ matrices, denoted as $$C^1_{l_q\times l_d \times 1}\cdots C^{l_g}_{l_q\times l_d \times 1}\;;$$
 \item \textbf{k-max pooling}: subsequently, the matching signals in \\$C^1,\cdots, C^{l_g}$ from these kernels 
 are further pooled with k-max pooling layers,
 keeping the top-$n_s$ strongest signals for each query term and CNN kernel pair,
 leading to $$P^1_{l_q\times n_s},\cdots, P^{l_g}_{l_q\times n_s}\;,$$
 which are further concatenated for individual query terms,
 resulting in a matrix $P_{l_q\times (l_g n_s)}$;
\newpage
 \item \textbf{combination of the signals from different query terms}: the signals in $P_{l_q\times (l_g n_s)}$, together with 
the inverse document frequency for individual query terms, 
are fed into a LSTM layer to generate the ultimate relevance score $\mathit{rel}(d, q)$.
\end{enumerate}

\textbf{Tweaks.}
Before moving on, we make two changes in order to ease
the development of the proposed model.
For simplicity, this revised model is denoted as PACRR in the following sections.
First, according to our pilot experiments, 
the performance of the model does not change
when replacing the LSTM layer with a stack of dense layers,
which have been demonstrated
to be able to simulate an arbitrary function~\cite{Goodfellow-et-al-2016}.
Such dense layers can easily be trained in parallel, 
leading to faster training~\cite{Goodfellow-et-al-2016},
whereas back-propagation through an LSTM layer 
is much more expensive due to its sequential nature.
From Section~\ref{sec.evaluation},
it can be seen that efficiency is important for this study
due to the number of model variants to be trained and 
the limited availability of hardware at our disposal.
Finally, another tweak is to switch the max-margin loss
to a cross-entropy loss as in Eq.~\ref{eq.crossentropylosss},
following~\cite{dehghani2017neural}, where it has been demonstrated that 
a cross-entropy loss may lead to better results.

\begin{equation}\label{eq.crossentropylosss}
 \mathcal{L}(q,d^+, d^-;\Theta)=-\mathit{log}\frac{\exp({\mathit{rel}(q,d^+)})}{\exp({\mathit{rel}(q,d^+)}) + \exp({\mathit{rel}(q,d^-)})}
\end{equation}


\section{Method} 
\label{sec.method}

In this section,
we describe the novel components in the Co-PACRR model as 
summarized in Figure~\ref{fig.model}.

\textbf{Disambiguation: checking local context when extracting matching signals}.
Beyond the query-document similarity matrix $\mathit{sim}_{l_q\times l_d}$ used by PACRR,
we introduce an input vector denoted as $\mathit{querysim}_{|d|}$ that
encodes the similarity between document context vectors and a query vector.
Document context vectors represent the meaning of text windows over the document,
while the query vector represents the query's meaning.
In particular,
the vector of a query $\mathit{queryvec}$ is computed by averaging the 
word vectors of all query terms.
Similarly,
given a position $i$ in a document,
its context vector of length, governed by $w_c$, is computed by averaging the embeddings of all the terms
appeared in its surrounding context,
$$\mathit{context2vec}(i) = \frac{\sum_{j\in [i-w_c, i+w_c]}\mathit{word2vec}(\mathbf{d}[i])}{2*w_c+1}\;.$$
Thereafter, the match between the query and a document context at position $i$
is computed by taking the cosine similarity between the query vector and context vector, that is,
$$\mathit{querysim}(i) = cosine(\mathit{context2vec}(i), \mathit{queryvec})\;.$$
We employ pre-trained
word2vec\footnote{\url{https://code.google.com/archive/p/word2vec/}}
embeddings due to their widespread availability.
In the future, one may desire to replace this with 
specialized embeddings such as dual embeddings~\cite{mitra2016dual}
or relevance-based embeddings~\cite{zamani2017relevance}.

Intuitively, to address the challenge of false positive matches 
stemming from ambiguity, 
the extracted matching signals on position $i$ are adjusted in the model
according to the corresponding similarity between its context and the query.
In particular, 
when combining the top-$n_s$ signals from individual query terms,
the corresponding similarities for these top-$n_s$ signals are also concatenated,
 making the matrices $P_{l_q\times (l_g n_s)}$ become $P_{l_q\times (2l_g n_s)}$. 
 This enables the aggregating layers, namely, a feed-forward 
network, to take any ambiguity into account
when determining the ultimate score.
For example, in the ``jaguar'' example from Section~\ref{sec.introduction},
if the context of ``jaguar'' consists of terms like ``big cat'' and ``habitat'',
the context will have a low similarity with a query context
containing terms such as ``SUV'' and ``price'',
informing the model that such occurrences of ``jaguar''
actually refer to a different concept than the one in the query.

\textbf{Cascade k-max pooling: encode the location of the relevance information.}
As discussed in Section~\ref{sec.introduction},
to put individual relevance signals into the context of the whole document,
both the strength and the positions of match signals matter. 
We propose to encode such global context by
conducting k-max pooling at multiple positions in a document, 
instead of pooling only on the entire document. 
For example, one could conduct multiple k-max pooling operations at
$25\%$, $50\%$,
$75\%$, and $100\%$ of a document, ending up with
$P_{l_q\times (4l_g n_s)}$. This corresponds to when a user 
sifts through a document and evaluates the gained useful information
after reading the first, second, third, and fourth quarters of the document.
The list of offsets at which cascade k-max pooling is conducted
is governed by an array $cpos$, e.g., $cpos=[25\%,50\%,75\%,100\%]$ in the above example.
We set the length of this array using a hyper-parameter $n_c$ and perform pooling at equal intervals.
For example, $n_c=4$ in the previous example, and $n_c=2$ results in $cpos=[50\%,100\%]$.

\textbf{Shuffling combination: regularizing the query-dependent information.}
As mentioned in Section~\ref{sec.introduction}, 
the combination of relevance signals among different query terms 
is supposed to be query-independent to avoid learning a dependency on query term positions.
In light of this, 
we propose to randomly shuffle rows in 
$P_{l_q\times (l_g n_s)}$ before aggregating them.
Note that each row contains signals for multiple n-gram lengths; shuffling the rows does not prevent the
model from recognizing n-grams.
We argue that, taking advantage of this independence, the shuffling 
regularizes the query-dependent information and
effectively improves the 
generalization ability of the model by making the computation of the relevance scores
depend solely on the importance of a query term ($\mathit{idf}$) and 
the relevance signals aggregated on it. 
This should be particularly 
helpful when training on short queries ($|q| < l_q$),
where padded zeros are normally in the tail of $\mathit{sim}_{l_q\times l_d}$~\cite{hui2017nrep}.
Without shuffling, a model might remember that the relevance signals 
at the tail of a query (i.e., the several final rows in $\mathit{sim}_{l_q\times l_d}$)
contribute very little and are mostly zero, leading to it mistakenly 
degrade the contribution from terms at tail positions when inferring relevance scores
for longer queries.


\section{Evaluation} 
\label{sec.evaluation}
In this section, we empirically compare the proposed 
Co-PACRR with multiple state-of-the-art neural IR models
using manual relevance judgments from six years of the \textsc{Trec} Web Track.
Following \cite{hui2017position}, the comparison is based on three benchmarks, namely,
re-ranking search results from a simple initial ranker, denoted as \textsc{RerankSimple},
re-ranking all runs from the \textsc{Trec} Web Track, denoted as \textsc{RerankALL},
and further examining the classification accuracy
in determining the order of document pairs, denoted as \textsc{PairAccuracy}.
We compare our model with multiple state-of-the-art neural IR models
including the PACRR model~\cite{hui2017position},
MatchPyramid~\cite{DBLP:journals/corr/PangLGXC16},
DRMM~\cite{guo2016deep},
the local model of DUET (DUETL)~\cite{mitra2017learning},
and the most recent K-NRM~\cite{xiong2017end} model.
As discussed in Section~\ref{sec.relatedwork}, 
our focus is on evaluating deep relevance matching models,
and hence the comparisons are limited to
1) modeling the interactions between a query and a document, 
excluding the learning-to-rank features for a single document or a query, e.g., PageRank scores, and 
2) modeling relevance matching rather than semantic matching~\cite{guo2016deep}.

\subsection{Experimental Setup}\label{sec.expsetting}
We rely on the 2009--2014 \textsc{Trec} Web
Track ad-hoc task benchmarks\footnote{\url{http://trec.nist.gov/tracks.html}}.
In total, there are 300 queries and around 100k judgments (qrels).
Six years (2009--14) of query-likelihood baselines (\textit{QL})  
 provided by the Lemur project's
 online Indri services\footnote{\url{http://boston.lti.cs.cmu.edu/Services/clueweb09_batch/}}
 \footnote{\url{http://boston.lti.cs.cmu.edu/Services/clueweb12_batch/}}
 serve as the initial ranker in \textsc{RerankSimple}. 
In addition, the search results from 
runs submitted by participants from each year are employed 
in the \textsc{RerankALL}, where
there are
71 (2009), 55 (2010),
62 (2011), 48 (2012), 50 (2013), and 27 (2014) runs.
ERR@20~\cite{Chapelle2009ERR} is
employed as evaluation measure, following 
the configuration in the \textsc{Trec} Web Track~\cite{collins2015trec}, which is computed with the script from 
\textsc{Trec}\footnote{http://trec.nist.gov/data/web/12/gdeval.pl}.
Note that ERR emphasizes the
quality of the top-ranked documents and heavily penalizes 
relevant documents 
that are ranked lower by a model when 
enough relevant documents have been observed earlier~\cite{Chapelle2009ERR}. 
This means that the improvement of the ERR for a model mainly comes from
improvements on queries for which search results at the top are not 
good enough from an initial ranker.

\paragraphHd{Training}
Models are trained and tested in a round-robin manner, 
using individual years as training, validation, 
and test data.
Specifically, the available judgments are considered 
in accordance with the individual years of the Web Track, with
50 queries per year.
Proceeding in a round-robin manner, 
we report test results on one year by 
using combinations of every four years 
and the two remaining years
for training and validation.
Model parameters and the number of training 
iterations are chosen by maximizing the ERR@20 on the validation set
for each training/validation combination separately.
Thereafter, the selected model is 
used to make predictions on the test data.
Hence, for each test year, 
there are five different predictions each from a training and validation
combination.
Akin to the procedure in cross-validation,
we report the average of these five test results as the ultimate results for individual test years,
and conduct a Student's t-test over them to determine whether there is a statistically significant difference
between different methods.
For example, a significant difference between two evaluated methods 
on a particular test year is claimed 
if there exists a significant difference between the two vectors with five scores 
for individual methods.
This was motivated by an observation that the closeness of the subsets for training and 
for validation can adversely influence the model selection. 
We argue that this approach minimizes
the effects of the choice of training and validation data.
Upper/lower-case characters
are employed to 
indicate the significant difference under
two-tailed Student's t-tests at 95\% or 90\% confidence levels 
relative to the corresponding approach,
denoted as 
P/p for PACRR, M/m for MatchPyramid,
D/d for DRMM,
L/l for DUETL and K/k for K-NRM.

\textbf{Variants of Co-PACRR.} 
With the proposed components, namely,
the cascade k-max pooling (C),
the disambiguation component (D), and 
the shuffling combination (S),
there are seven model variants in total by including or 
excluding one of the three building blocks.
They are denoted as X(XX)-PACRR, where the X represents the 
building blocks that are turned on. For example, 
with cascade k-max pooling and shuffling combination turned on, the model is denoted
as CS-PACRR. Meanwhile, with all three components,
namely CDS-PACRR, the model is simply referred to as Co-PACRR.
In evaluations based on the \textsc{RerankSimple} and 
\textsc{RerankALL} benchmarks,
only the results for Co-PACRR are reported. Meanwhile, the 
results for the other six variants are reported in Section~\ref{sec.analysis}
on the \textsc{PairAccuracy} benchmark
for an ablation test.

\textbf{Choice of hyper-parameters.}
In this work, we focus on evaluating the effects of the proposed 
building blocks and their interactions, 
without exhaustively fine-tuning hyper-parameters due to limited computing resources.
For the disambiguation building block, 
we fix the size of the context window as $w_c=4$ on both sides, 
leading to a context vector computed over 9 terms, namely, $4+4+1$.
For the cascade component, we conduct k-max pooling
with $cpos=[25\%, 50\%, 75\%, 100\%]$, namely, $n_c=4$.
For the combination phase, we use two fully connected layers of size 16.
Apart from the two modifications mentioned in Section~\ref{sec.background}, 
we further 
fix the model choices for PACRR following the original configurations~\cite{hui2017position}.
In particular, the PACRR-firstk variant is employed,
fixing the unified similarity matrix dimensions $l_d=800$ and $l_q=16$,
the k-max pooling size $n_s=3$, the maximum n-gram size $l_g=3$, and the number of 
filters used in convolutional layers is $n_f=32$. 
Beyond that, we fix the batch size to $16$ and we 
train individual models to at most $150$ iterations.
Note that most of the aforementioned hyper-parameters can be tuned given
sufficient time and hardware, and the chosen parameters follow
those in Hui et al.~\cite{hui2017position} or are based on preliminary
experiments for a better focus on the proposed models.
In Section~\ref{sec.tuning} we consider the impact of the disambiguation parameter $w_c$
and the cascade parameter $n_c$.

Due to the availability of training data,
K-NRM is trained   
 with a frozen word embedding layer,
 and with an extra fully connected middle layer including 30 neurons to
 partially compensate for lost strength due to the frozen word embeddings.
 This is slightly different from the model architecture described
 in Xiong et al.~\cite{xiong2017end}.
 This setting also serves for the purpose of  
 allowing fair model comparisons, given that all the compared models
 could be co-trained with  
 the word embeddings, resulting in a better model capacity
 at the costs of prolonged training times and a need for much more training data~\cite{hui2017position}.
 Note that with the frozen embedding layer, the evaluation can
 focus on the model strength that comes from different model 
 architectures, demonstrating the capacity of relatively small models in performing
 ad-hoc retrieval.
 All the models are trained with a cross-entropy loss as summarized in
Eq.~\ref{eq.crossentropylosss}, given that different loss functions 
can also influence the results.

\subsection{Results for Co-PACRR}
\begin{table*}[!ht]
  \caption{ERR@20 on \textsc{Trec} Web Track 2009--14
 when re-ranking search results from \textit{QL}. 
The relative improvements (\%) relative to QL 
and ranks among all runs within the respective years
 according to ERR@20 are also reported.}\label{tab.neup.rrql}
\centering
\ra
\resizebox{0.9\textwidth}{!}{
\begin{tabular}{@{}l|rrrrrr@{}}
\toprule
Year&Co-PACRR&PACRR&MatchPyramid&DRMM&DUETL&K-NRM\\
\midrule
wt09&0.096 (D$\uparrow$) \phantom{0}6\% 47&0.102 (D$\uparrow$) \phantom{0}13\% 41&0.103 \phantom{0}14\% 38&0.086 (P$\downarrow$) \phantom{0}-5\% 50&0.092 \phantom{0}1\% 45&0.091 \phantom{0}1\% 48\\
wt10&0.160 (P$\uparrow$K$\uparrow$D$\uparrow$L$\uparrow$M$\uparrow$) \phantom{0}136\% 3&0.146 (K$\uparrow$D$\uparrow$L$\uparrow$m$\uparrow$) \phantom{0}116\% 4&0.131 (p$\downarrow$L$\uparrow$) \phantom{0}93\% 9&0.131 (P$\downarrow$L$\uparrow$) \phantom{0}92\% 9&0.103 (P$\downarrow$K$\downarrow$D$\downarrow$M$\downarrow$) \phantom{0}52\% 25&0.128 (P$\downarrow$L$\uparrow$) \phantom{0}88\% 10\\
wt11&0.167 (P$\uparrow$K$\uparrow$D$\uparrow$L$\uparrow$M$\uparrow$) \phantom{0}52\% 2&0.139 (k$\uparrow$L$\uparrow$M$\uparrow$) \phantom{0}26\% 15&0.114 (P$\downarrow$K$\downarrow$D$\downarrow$) \phantom{0}3\% 31&0.133 (L$\uparrow$M$\uparrow$) \phantom{0}21\% 19&0.112 (P$\downarrow$K$\downarrow$D$\downarrow$) \phantom{0}1\% 35&0.129 (p$\downarrow$L$\uparrow$M$\uparrow$) \phantom{0}17\% 23\\
wt12&0.359 (K$\uparrow$D$\uparrow$L$\uparrow$M$\uparrow$) \phantom{0}99\% 1&0.363 (K$\uparrow$D$\uparrow$L$\uparrow$M$\uparrow$) \phantom{0}101\% 1&0.244 (P$\downarrow$D$\downarrow$) \phantom{0}35\% 15&0.320 (P$\downarrow$K$\uparrow$L$\uparrow$M$\uparrow$) \phantom{0}77\% 3&0.206 (P$\downarrow$K$\downarrow$D$\downarrow$) \phantom{0}13\% 22&0.269 (P$\downarrow$D$\downarrow$L$\uparrow$) \phantom{0}49\% 11\\
wt13&0.189 (K$\uparrow$D$\uparrow$L$\uparrow$M$\uparrow$) \phantom{0}82\% 1&0.184 (K$\uparrow$D$\uparrow$L$\uparrow$M$\uparrow$) \phantom{0}77\% 1&0.131 (P$\downarrow$D$\downarrow$) \phantom{0}26\% 18&0.166 (P$\downarrow$K$\uparrow$L$\uparrow$M$\uparrow$) \phantom{0}60\% 3&0.130 (P$\downarrow$D$\downarrow$) \phantom{0}25\% 20&0.141 (P$\downarrow$D$\downarrow$) \phantom{0}36\% 12\\
wt14&0.232 (P$\uparrow$K$\uparrow$D$\uparrow$L$\uparrow$M$\uparrow$) \phantom{0}84\% 1&0.210 (K$\uparrow$d$\uparrow$L$\uparrow$M$\uparrow$) \phantom{0}67\% 4&0.163 (P$\downarrow$D$\downarrow$) \phantom{0}29\% 19&0.191 (p$\downarrow$K$\uparrow$L$\uparrow$M$\uparrow$) \phantom{0}52\% 10&0.159 (P$\downarrow$D$\downarrow$) \phantom{0}26\% 20&0.167 (P$\downarrow$D$\downarrow$) \phantom{0}32\% 17\\
\bottomrule
\end{tabular}
}
\end{table*}

\textbf{\textsc{RerankSimple}}.
We first examine how well the proposed model performs 
when re-ranking search results from a simple initial ranker, namely,
the query-likelihood (QL) model,
to put our results in context as in Guo et al.~\cite{guo2016deep}.
The ultimate quality of the re-ranked search results
depends on both the strength of the initial ranker and the quality of the re-ranker.
The query-likelihood model, as one of the most widely used retrieval
models, is used due to its efficiency and practical availability,
given that it is included in most retrieval toolkits like Terrier~\cite{ounis06terrier}. 
The results are summarized in Table~\ref{tab.neup.rrql}. 
The ERR@20 of the re-ranked runs is reported, together with the 
improvements relative to the original QL.
The ranks of the re-ranked runs are also reported when sorting the re-ranked search results
together with other competing runs from the same year according to ERR@20.

It can be seen that, by simply re-ranking the search results 
from the query-likelihood method, 
Co-PACRR
can already achieve the top-3 best results
in 2010--14.
Whereas for 2009, very limited improvements are observed.
Combined with Table~\ref{tab.neup.rrallcomp},
though variants of Co-PACRR can improve different runs in \textsc{Trec}
around 90\%, the relative improvements w.r.t. QL are less than 10\%,
which is worse than the improvements from PACRR and MatchPyramid on 2009. 
This illustrates that the re-ranking model cannot work independently,
as its performance highly depends on the initial ranker.
Actually, in Table~\ref{tab.neup.rrql} all compared models experience difficulties 
in improving QL on 2009, where DRMM even receives a worse ranking.
This might be partially explained by the difference of the initial ranker
in terms of the recall rate.
Intuitively, 
there should be enough relevant documents to be re-ranked in 
the initial ranking, otherwise the re-ranker is unable to achieve anything, no matter its quality.
The recall rates of QL in different years are as follows (in parentheses):
2009 (0.35), 2010 (0.37), 2011 (0.67), 2012 (0.46), 2013 (0.61),
and 2014 (0.68), where 2009 witnesses the lowest recall.
However, there may also be other causes for these results.

\textbf{\textsc{RerankALL}}.
Given that the search results from QL
only account for a small subset of all judged documents, and,
more importantly, that
the performance of a re-ranker also depends on the initial runs,
we evaluate our re-ranker's performance by re-ranking all
submitted runs from the \textsc{Trec} Web Track 2009--14. 
This evaluation focuses on two aspects:
how many different runs we can improve upon and by how much we improve.
The former aspect is about the \textit{adaptability} of a neural IR model,
investigating whether it can make improvements based on different kinds
of retrieval models, while the latter aspect focuses on 
the \textit{magnitude of improvements}.
Table~\ref{tab.neup.rrallrank} summarizes the percentages of systems 
that see improvements based on ERR@20 out of the total number 
of systems in each year.
In Table~\ref{tab.neup.rrallcomp}, we further report the 
average percentage of improvements.

Table~\ref{tab.neup.rrallrank} demonstrates that 
at least 90\% of runs, and on average more than 96\% of runs, can be improved by 
Co-PACRR, which implies a good adaptability, namely, the proposed
Co-PACRR can work together with a wide range of initial rankers using different methods.
Compared with other neural IR models, 
in terms of the absolute numbers, 
Co-PACRR improves the highest number of systems in all the years;
when conducting significance tests,
in three out of six years, the proposed Co-PACRR significantly 
outperforms all the baselines.
Noticeably, Co-PACRR uniformly achieves good results on all six years, whereas 
all other methods fail to improve more than 75\% of systems in at least one year.
There are similar observations for the average improvements 
shown in Table~\ref{tab.neup.rrallcomp},
where Co-PACRR performs best in terms of the average improvements 
for all six years; on four out of six years Co-PACRR leads other methods with
a significant difference.
This table shows that Co-PACRR can improve different runs in each year 
by at least 34\% on average.

\begin{table*}[!ht]
  \caption{ 
The percentage of runs that show improvements in terms of ERR@20 
when re-ranking all runs from the \textsc{Trec} Web Track 2009--14.}\label{tab.neup.rrallrank}
\centering
\ra
\resizebox{0.7\textwidth}{!}{
\begin{tabular}{l|rrrrrr}
\toprule
Year&Co-PACRR&PACRR&MatchPyramid&DRMM&DUETL&K-NRM\\
\midrule
wt09&\phantom{0}90\% (D$\uparrow$L$\uparrow$)&\phantom{0}93\% (D$\uparrow$L$\uparrow$)&\phantom{0}88\% (D$\uparrow$l$\uparrow$)&\phantom{0}70\% (P$\downarrow$K$\downarrow$M$\downarrow$)&\phantom{0}74\% (P$\downarrow$K$\downarrow$m$\downarrow$)&\phantom{0}89\% (D$\uparrow$L$\uparrow$)\\
wt10&\phantom{0}98\% (K$\uparrow$D$\uparrow$L$\uparrow$M$\uparrow$)&\phantom{0}96\% (D$\uparrow$L$\uparrow$M$\uparrow$)&\phantom{0}89\% (P$\downarrow$K$\downarrow$L$\uparrow$)&\phantom{0}91\% (P$\downarrow$K$\downarrow$L$\uparrow$)&\phantom{0}74\% (P$\downarrow$K$\downarrow$D$\downarrow$M$\downarrow$)&\phantom{0}95\% (D$\uparrow$L$\uparrow$M$\uparrow$)\\
wt11&\phantom{0}98\% (P$\uparrow$K$\uparrow$D$\uparrow$L$\uparrow$M$\uparrow$)&\phantom{0}71\% (D$\uparrow$L$\uparrow$M$\uparrow$)&\phantom{0}15\% (P$\downarrow$K$\downarrow$D$\downarrow$L$\downarrow$)&\phantom{0}42\% (P$\downarrow$K$\downarrow$L$\uparrow$M$\uparrow$)&\phantom{0}21\% (P$\downarrow$K$\downarrow$D$\downarrow$M$\uparrow$)&\phantom{0}69\% (D$\uparrow$L$\uparrow$M$\uparrow$)\\
wt12&\phantom{0}98\% (P$\uparrow$K$\uparrow$d$\uparrow$L$\uparrow$M$\uparrow$)&\phantom{0}95\% (K$\uparrow$L$\uparrow$M$\uparrow$)&\phantom{0}73\% (P$\downarrow$k$\downarrow$D$\downarrow$)&\phantom{0}94\% (K$\uparrow$L$\uparrow$M$\uparrow$)&\phantom{0}72\% (P$\downarrow$k$\downarrow$D$\downarrow$)&\phantom{0}83\% (P$\downarrow$D$\downarrow$l$\uparrow$m$\uparrow$)\\
wt13&\phantom{0}93\% (p$\uparrow$K$\uparrow$d$\uparrow$L$\uparrow$M$\uparrow$)&\phantom{0}86\% (K$\uparrow$L$\uparrow$M$\uparrow$)&\phantom{0}56\% (P$\downarrow$D$\downarrow$)&\phantom{0}87\% (K$\uparrow$L$\uparrow$M$\uparrow$)&\phantom{0}43\% (P$\downarrow$K$\downarrow$D$\downarrow$)&\phantom{0}63\% (P$\downarrow$D$\downarrow$L$\uparrow$)\\
wt14&\phantom{0}96\% (K$\uparrow$D$\uparrow$L$\uparrow$M$\uparrow$)&\phantom{0}84\% (K$\uparrow$L$\uparrow$M$\uparrow$)&\phantom{0}61\% (P$\downarrow$K$\uparrow$L$\uparrow$)&\phantom{0}69\% (K$\uparrow$L$\uparrow$)&\phantom{0}39\% (P$\downarrow$D$\downarrow$M$\downarrow$)&\phantom{0}43\% (P$\downarrow$D$\downarrow$M$\downarrow$)\\
\bottomrule
\end{tabular}
}
\end{table*}

\begin{table*}[!t]
  \caption{
 The average differences of the measure score for individual runs
 when re-ranking all runs from the \textsc{Trec} Web Track 2009--14
  based on ERR@20.}\label{tab.neup.rrallcomp}
\centering
\ra
\resizebox{0.7\textwidth}{!}{
\begin{tabular}{@{}l|rrrrrr@{}}
\toprule
Year&Co-PACRR&PACRR&MatchPyramid&DRMM&DUETL&K-NRM\\
\midrule
wt09&\phantom{0}47\% (p$\uparrow$K$\uparrow$D$\uparrow$L$\uparrow$M$\uparrow$)&\phantom{0}42\% (K$\uparrow$D$\uparrow$L$\uparrow$M$\uparrow$)&\phantom{0}29\% (P$\downarrow$D$\uparrow$L$\uparrow$)&\phantom{0}17\% (P$\downarrow$K$\downarrow$M$\downarrow$)&\phantom{0}16\% (P$\downarrow$K$\downarrow$M$\downarrow$)&\phantom{0}35\% (P$\downarrow$D$\uparrow$L$\uparrow$)\\
wt10&\phantom{0}93\% (P$\uparrow$K$\uparrow$D$\uparrow$L$\uparrow$M$\uparrow$)&\phantom{0}76\% (D$\uparrow$L$\uparrow$M$\uparrow$)&\phantom{0}51\% (P$\downarrow$K$\downarrow$L$\uparrow$)&\phantom{0}48\% (P$\downarrow$K$\downarrow$L$\uparrow$)&\phantom{0}27\% (P$\downarrow$K$\downarrow$D$\downarrow$M$\downarrow$)&\phantom{0}68\% (D$\uparrow$L$\uparrow$M$\uparrow$)\\
wt11&\phantom{0}39\% (P$\uparrow$K$\uparrow$D$\uparrow$L$\uparrow$M$\uparrow$)&\phantom{0}10\% (D$\uparrow$L$\uparrow$M$\uparrow$)&\phantom{0}-22\% (P$\downarrow$K$\downarrow$D$\downarrow$l$\downarrow$)&\phantom{0}-3\% (P$\downarrow$K$\downarrow$L$\uparrow$M$\uparrow$)&\phantom{0}-17\% (P$\downarrow$K$\downarrow$D$\downarrow$m$\uparrow$)&\phantom{0}8\% (D$\uparrow$L$\uparrow$M$\uparrow$)\\
wt12&\phantom{0}84\% (K$\uparrow$D$\uparrow$L$\uparrow$M$\uparrow$)&\phantom{0}74\% (K$\uparrow$L$\uparrow$M$\uparrow$)&\phantom{0}28\% (P$\downarrow$D$\downarrow$)&\phantom{0}69\% (K$\uparrow$L$\uparrow$M$\uparrow$)&\phantom{0}29\% (P$\downarrow$D$\downarrow$)&\phantom{0}44\% (P$\downarrow$D$\downarrow$)\\
wt13&\phantom{0}38\% (K$\uparrow$D$\uparrow$L$\uparrow$M$\uparrow$)&\phantom{0}30\% (K$\uparrow$L$\uparrow$M$\uparrow$)&\phantom{0}4\% (P$\downarrow$D$\downarrow$)&\phantom{0}22\% (K$\uparrow$L$\uparrow$M$\uparrow$)&\phantom{0}-4\% (P$\downarrow$K$\downarrow$D$\downarrow$)&\phantom{0}11\% (P$\downarrow$D$\downarrow$L$\uparrow$)\\
wt14&\phantom{0}34\% (P$\uparrow$K$\uparrow$D$\uparrow$L$\uparrow$M$\uparrow$)&\phantom{0}20\% (K$\uparrow$d$\uparrow$L$\uparrow$M$\uparrow$)&\phantom{0}6\% (P$\downarrow$K$\uparrow$L$\uparrow$)&\phantom{0}10\% (p$\downarrow$K$\uparrow$L$\uparrow$)&\phantom{0}-4\% (P$\downarrow$D$\downarrow$M$\downarrow$)&\phantom{0}-4\% (P$\downarrow$D$\downarrow$M$\downarrow$)\\
\bottomrule
\end{tabular}
}
\end{table*}

\textbf{\textsc{PairAccuracy}}.
Ideally, a re-ranking model should 
make correct decisions when ranking all kinds of documents. 
Therefore, we further rely on a pairwise ranking task 
to compare different models in this regard. 
Compared with the other two benchmarks, 
we argue that \textsc{PairAccuracy} can lead to more comprehensive
and more robust comparisons, as a result of its inclusion of all the labeled 
ground-truth data and its removal of the effects of initial rankers.

In particular, given a query and a set of documents, 
different models assign a score to each document
according to their inferred relevance relative to the given query.
Thereafter, all pairs of documents are examined and
the pairs that are ranked in concordance 
with the ground-truth judgments from \textsc{Trec}
are deemed correct, based on which 
an aggregated accuracy is reported on all such document pairs
in different years.
For example, given query $q$ and two documents $d_1$ and $d_2$,
along with their ground-truth judgments $\mathit{label}(d_1)$ 
and $\mathit{label}(d_2)$, 
a re-ranking model provides their relevance scores 
as $\mathit{rel}(q, d_1)$ and  $\mathit{rel}(q, d_2)$.
The re-ranking model is correct when it predicts
these two documents to be ranked in the same order as in the ranking
from the ground-truth label, e.g.,
$\mathit{rel}(q, d_1)>\mathit{rel}(q, d_2)$ and
$\mathit{label}(d_1)> \mathit{label}(d_2)$.
The relevance judgments in the \textsc{Trec} Web Track
include up to six relevance levels: junk pages (Junk),
non-relevant (NRel), relevant (Rel), highly relevant (HRel), key pages (Key), 
and navigational pages (Nav).
Note that the label Nav actually indicates 
that a document can satisfy
a navigational intent
rather than assigning a degree of relevance as Rel and HRel, 
which makes it difficult to compare navigational documents with other kinds of 
relevant documents,
e.g., a navigational document versus a document labeled as HRel.
Thus, documents labeled with Nav are not considered in this task.
Moreover, documents labeled as Junk and NRel,
and
documents labeled as HRel and Key are 
merged into NRel and HRel, respectively, due to 
their limited number.
After aggregating the labels as described,
all pairs of documents with different labels are generated as test pairs.
From the ``volume'' and ``\# queries'' columns in Table~\ref{tab.neup.docpair}, 
we can see that
different label pairs actually account for quite different volumes in the ground truth,
making their respective degrees of influence different.
On the other hand, 
different label pairs actually also represent 
different difficulties in making a correct prediction,
as the closeness of two documents in terms of 
their relevance determines the difficulty of the predictions.
Intuitively, it is easier to distinguish between HRel and NRel documents
than to compare a HRel document with a Rel document.
Actually, human assessors tend to also disagree more when dealing 
with document pairs that are very close with each  other in terms 
of their relevance~\cite{alonso2012using}.
It can also be seen that
these three label pairs being considered 
account for 95\% of
all document pairs from Table~\ref{tab.neup.docpair}.

From the upper part of Table~\ref{tab.neup.docpair},
for the label pair HRel-NRel,
Co-PACRR achieves the highest accuracy in terms of the absolute number, and 
significantly outperforms all baselines on three years.
We have similar observations for Rel-NRel,
where, however, Co-PACRR performs worse than PACRR in 2014.
As for the label pair HRel-Rel, however,
Co-PACRR performs very close to the other models, and on 2011,
it performs worse than DUETL.
Therefore, we can conclude that Co-PACRR 
outperforms the other baseline results when comparing  
documents that are far away in terms of relevance, while
performing similarly in dealing with harder pairs.
In terms of the absolute accuracy, 
on average, Co-PACRR yields correct predictions on
78.7\%, 73.6\%, and 58.7\% of document pairs 
for the label pairs HRel--NRel, Rel--NRel, and HRel--Rel,
respectively, where the decreasing 
accuracy confirms the different difficulties in making predictions for
different kinds of pairs.

\begin{table*}[!ht]
  \caption{Comparisons among tested methods in terms of 
		accuracy in ranking document pairs with different label pairs. 
		The columns ``volume'' and ``\#~queries''
		record the occurrences of each label combination out of the total pairs,
		and the number of queries that include
		a particular label combination among all six years, respectively. }\label{tab.neup.docpair}
\centering
\ra
\resizebox{0.9\textwidth}{!}{
\begin{tabular}{@{}llll|rrrrrr@{}}
\toprule
Label Pair&volume~(\%)&\# queries&Year&Co-PACRR&PACRR&MatchPyramid&DRMM&DUETL&K-NRM\\
\midrule
\multirow{6}{*}{\textit{HRel}-\textit{NRel}}&
\multirow{6}{*}{23.1\%}&
\multirow{6}{*}{262}
&wt09&\phantom{0}0.720 (P$\uparrow$K$\uparrow$D$\uparrow$L$\uparrow$M$\uparrow$)&\phantom{0}0.695 (D$\uparrow$L$\uparrow$M$\uparrow$)&\phantom{0}0.654 (P$\downarrow$K$\downarrow$D$\uparrow$L$\uparrow$)&\phantom{0}0.597 (P$\downarrow$K$\downarrow$M$\downarrow$)&\phantom{0}0.593 (P$\downarrow$K$\downarrow$M$\downarrow$)&\phantom{0}0.689 (D$\uparrow$L$\uparrow$M$\uparrow$)\\
&&&wt10&\phantom{0}0.850 (P$\uparrow$K$\uparrow$D$\uparrow$L$\uparrow$M$\uparrow$)&\phantom{0}0.831 (k$\uparrow$D$\uparrow$L$\uparrow$M$\uparrow$)&\phantom{0}0.768 (P$\downarrow$D$\uparrow$L$\uparrow$)&\phantom{0}0.740 (P$\downarrow$K$\downarrow$L$\uparrow$M$\downarrow$)&\phantom{0}0.677 (P$\downarrow$K$\downarrow$D$\downarrow$M$\downarrow$)&\phantom{0}0.797 (p$\downarrow$D$\uparrow$L$\uparrow$)\\
&&&wt11&\phantom{0}0.829 (P$\uparrow$K$\uparrow$D$\uparrow$L$\uparrow$M$\uparrow$)&\phantom{0}0.778 (K$\uparrow$D$\uparrow$L$\uparrow$M$\uparrow$)&\phantom{0}0.693 (P$\downarrow$K$\downarrow$D$\downarrow$L$\uparrow$)&\phantom{0}0.728 (P$\downarrow$k$\downarrow$L$\uparrow$M$\uparrow$)&\phantom{0}0.638 (P$\downarrow$K$\downarrow$D$\downarrow$M$\downarrow$)&\phantom{0}0.749 (P$\downarrow$d$\uparrow$L$\uparrow$M$\uparrow$)\\
&&&wt12&\phantom{0}0.801 (K$\uparrow$D$\uparrow$L$\uparrow$M$\uparrow$)&\phantom{0}0.790 (K$\uparrow$D$\uparrow$L$\uparrow$M$\uparrow$)&\phantom{0}0.703 (P$\downarrow$)&\phantom{0}0.672 (P$\downarrow$K$\downarrow$)&\phantom{0}0.683 (P$\downarrow$)&\phantom{0}0.728 (P$\downarrow$D$\uparrow$)\\
&&&wt13&\phantom{0}0.752 (K$\uparrow$D$\uparrow$L$\uparrow$M$\uparrow$)&\phantom{0}0.744 (K$\uparrow$D$\uparrow$L$\uparrow$M$\uparrow$)&\phantom{0}0.654 (P$\downarrow$L$\uparrow$)&\phantom{0}0.648 (P$\downarrow$l$\uparrow$)&\phantom{0}0.636 (P$\downarrow$K$\downarrow$d$\downarrow$M$\downarrow$)&\phantom{0}0.663 (P$\downarrow$L$\uparrow$)\\
&&&wt14&\phantom{0}0.772 (K$\uparrow$D$\uparrow$L$\uparrow$M$\uparrow$)&\phantom{0}0.770 (K$\uparrow$D$\uparrow$L$\uparrow$M$\uparrow$)&\phantom{0}0.670 (P$\downarrow$K$\uparrow$D$\uparrow$L$\uparrow$)&\phantom{0}0.653 (P$\downarrow$M$\downarrow$)&\phantom{0}0.639 (P$\downarrow$M$\downarrow$)&\phantom{0}0.640 (P$\downarrow$M$\downarrow$)\\
\midrule
\multirow{6}{*}{\textit{HRel}-\textit{Rel}}&
\multirow{6}{*}{\phantom{0}8.4\%}&
\multirow{6}{*}{257}
&wt09&\phantom{0}0.545 (L$\uparrow$)&\phantom{0}0.534&\phantom{0}0.537&\phantom{0}0.543 (L$\uparrow$)&\phantom{0}0.529 (K$\downarrow$D$\downarrow$)&\phantom{0}0.542 (L$\uparrow$)\\
&&&wt10&\phantom{0}0.576 (D$\uparrow$L$\uparrow$)&\phantom{0}0.577 (D$\uparrow$L$\uparrow$)&\phantom{0}0.591 (D$\uparrow$L$\uparrow$)&\phantom{0}0.542 (P$\downarrow$M$\downarrow$)&\phantom{0}0.545 (P$\downarrow$M$\downarrow$)&\phantom{0}0.572\\
&&&wt11&\phantom{0}0.576 (P$\uparrow$K$\uparrow$D$\downarrow$L$\uparrow$m$\downarrow$)&\phantom{0}0.522 (D$\downarrow$M$\downarrow$)&\phantom{0}0.589 (P$\uparrow$K$\uparrow$D$\downarrow$L$\uparrow$)&\phantom{0}0.615 (P$\uparrow$K$\uparrow$L$\uparrow$M$\uparrow$)&\phantom{0}0.507 (D$\downarrow$M$\downarrow$)&\phantom{0}0.518 (D$\downarrow$M$\downarrow$)\\
&&&wt12&\phantom{0}0.645 (K$\uparrow$D$\uparrow$L$\uparrow$M$\uparrow$)&\phantom{0}0.644 (K$\uparrow$D$\uparrow$L$\uparrow$M$\uparrow$)&\phantom{0}0.575 (P$\downarrow$D$\uparrow$)&\phantom{0}0.528 (P$\downarrow$K$\downarrow$L$\downarrow$M$\downarrow$)&\phantom{0}0.583 (P$\downarrow$D$\uparrow$)&\phantom{0}0.583 (P$\downarrow$D$\uparrow$)\\
&&&wt13&\phantom{0}0.575 (K$\uparrow$D$\uparrow$L$\uparrow$M$\uparrow$)&\phantom{0}0.579 (m$\uparrow$)&\phantom{0}0.551 (p$\downarrow$)&\phantom{0}0.560&\phantom{0}0.558&\phantom{0}0.551\\
&&&wt14&\phantom{0}0.602 (P$\uparrow$K$\uparrow$D$\uparrow$L$\uparrow$M$\uparrow$)&\phantom{0}0.575 (K$\uparrow$d$\uparrow$)&\phantom{0}0.569 (K$\uparrow$D$\uparrow$)&\phantom{0}0.558 (p$\downarrow$K$\uparrow$M$\downarrow$)&\phantom{0}0.560 (K$\uparrow$)&\phantom{0}0.507 (P$\downarrow$D$\downarrow$L$\downarrow$M$\downarrow$)\\
\midrule
\multirow{6}{*}{\textit{Rel}-\textit{NRel}}&
\multirow{6}{*}{63.5\%}&
\multirow{6}{*}{290}
&wt09&\phantom{0}0.676 (P$\uparrow$K$\uparrow$D$\uparrow$L$\uparrow$M$\uparrow$)&\phantom{0}0.663 (K$\uparrow$D$\uparrow$L$\uparrow$M$\uparrow$)&\phantom{0}0.619 (P$\downarrow$K$\downarrow$D$\uparrow$L$\uparrow$)&\phantom{0}0.555 (P$\downarrow$K$\downarrow$M$\downarrow$)&\phantom{0}0.563 (P$\downarrow$K$\downarrow$M$\downarrow$)&\phantom{0}0.650 (P$\downarrow$D$\uparrow$L$\uparrow$M$\uparrow$)\\
&&&wt10&\phantom{0}0.811 (P$\uparrow$K$\uparrow$D$\uparrow$L$\uparrow$M$\uparrow$)&\phantom{0}0.791 (K$\uparrow$D$\uparrow$L$\uparrow$M$\uparrow$)&\phantom{0}0.708 (P$\downarrow$K$\downarrow$L$\uparrow$)&\phantom{0}0.710 (P$\downarrow$K$\downarrow$L$\uparrow$)&\phantom{0}0.639 (P$\downarrow$K$\downarrow$D$\downarrow$M$\downarrow$)&\phantom{0}0.751 (P$\downarrow$D$\uparrow$L$\uparrow$M$\uparrow$)\\
&&&wt11&\phantom{0}0.787 (P$\uparrow$K$\uparrow$D$\uparrow$L$\uparrow$M$\uparrow$)&\phantom{0}0.770 (K$\uparrow$D$\uparrow$L$\uparrow$M$\uparrow$)&\phantom{0}0.616 (P$\downarrow$K$\downarrow$)&\phantom{0}0.607 (P$\downarrow$K$\downarrow$)&\phantom{0}0.621 (P$\downarrow$K$\downarrow$)&\phantom{0}0.711 (P$\downarrow$D$\uparrow$L$\uparrow$M$\uparrow$)\\
&&&wt12&\phantom{0}0.735 (p$\uparrow$K$\uparrow$D$\uparrow$L$\uparrow$M$\uparrow$)&\phantom{0}0.721 (K$\uparrow$D$\uparrow$L$\uparrow$M$\uparrow$)&\phantom{0}0.640 (P$\downarrow$K$\downarrow$L$\uparrow$)&\phantom{0}0.651 (P$\downarrow$k$\downarrow$L$\uparrow$)&\phantom{0}0.616 (P$\downarrow$K$\downarrow$D$\downarrow$M$\downarrow$)&\phantom{0}0.673 (P$\downarrow$d$\uparrow$L$\uparrow$M$\uparrow$)\\
&&&wt13&\phantom{0}0.700 (K$\uparrow$D$\uparrow$L$\uparrow$M$\uparrow$)&\phantom{0}0.689 (K$\uparrow$D$\uparrow$L$\uparrow$M$\uparrow$)&\phantom{0}0.612 (P$\downarrow$D$\uparrow$L$\uparrow$)&\phantom{0}0.589 (P$\downarrow$K$\downarrow$M$\downarrow$)&\phantom{0}0.579 (P$\downarrow$K$\downarrow$M$\downarrow$)&\phantom{0}0.623 (P$\downarrow$D$\uparrow$L$\uparrow$)\\
&&&wt14&\phantom{0}0.708 (p$\downarrow$K$\uparrow$D$\uparrow$L$\uparrow$M$\uparrow$)&\phantom{0}0.717 (K$\uparrow$D$\uparrow$L$\uparrow$M$\uparrow$)&\phantom{0}0.620 (P$\downarrow$K$\downarrow$D$\uparrow$L$\uparrow$)&\phantom{0}0.597 (P$\downarrow$K$\downarrow$M$\downarrow$)&\phantom{0}0.586 (P$\downarrow$K$\downarrow$M$\downarrow$)&\phantom{0}0.647 (P$\downarrow$D$\uparrow$L$\uparrow$M$\uparrow$)\\
\bottomrule

Label Pair&volume~(\%)&\# queries&Year&C-PACRR&D-PACRR&S-PACRR&CD-PACRR&CS-PACRR&DS-PACRR\\
\midrule
\multirow{6}{*}{\textit{HRel}-\textit{NRel}}&
\multirow{6}{*}{23.1\%}&
\multirow{6}{*}{262}
&wt09&\phantom{0}0.702 (K$\uparrow$D$\uparrow$L$\uparrow$M$\uparrow$)&\phantom{0}0.701 (K$\uparrow$D$\uparrow$L$\uparrow$M$\uparrow$)&\phantom{0}0.704 (p$\uparrow$K$\uparrow$D$\uparrow$L$\uparrow$M$\uparrow$)&\phantom{0}0.704 (K$\uparrow$D$\uparrow$L$\uparrow$M$\uparrow$)&\phantom{0}0.716 (P$\uparrow$K$\uparrow$D$\uparrow$L$\uparrow$M$\uparrow$)&\phantom{0}0.713 (P$\uparrow$K$\uparrow$D$\uparrow$L$\uparrow$M$\uparrow$)\\
&&&wt10&\phantom{0}0.839 (p$\uparrow$K$\uparrow$D$\uparrow$L$\uparrow$M$\uparrow$)&\phantom{0}0.842 (P$\uparrow$K$\uparrow$D$\uparrow$L$\uparrow$M$\uparrow$)&\phantom{0}0.843 (P$\uparrow$K$\uparrow$D$\uparrow$L$\uparrow$M$\uparrow$)&\phantom{0}0.842 (p$\uparrow$K$\uparrow$D$\uparrow$L$\uparrow$M$\uparrow$)&\phantom{0}0.848 (P$\uparrow$K$\uparrow$D$\uparrow$L$\uparrow$M$\uparrow$)&\phantom{0}0.843 (p$\uparrow$K$\uparrow$D$\uparrow$L$\uparrow$M$\uparrow$)\\
&&&wt11&\phantom{0}0.808 (K$\uparrow$D$\uparrow$L$\uparrow$M$\uparrow$)&\phantom{0}0.820 (P$\uparrow$K$\uparrow$D$\uparrow$L$\uparrow$M$\uparrow$)&\phantom{0}0.824 (P$\uparrow$K$\uparrow$D$\uparrow$L$\uparrow$M$\uparrow$)&\phantom{0}0.810 (p$\uparrow$K$\uparrow$D$\uparrow$L$\uparrow$M$\uparrow$)&\phantom{0}0.821 (P$\uparrow$K$\uparrow$D$\uparrow$L$\uparrow$M$\uparrow$)&\phantom{0}0.836 (P$\uparrow$K$\uparrow$D$\uparrow$L$\uparrow$M$\uparrow$)\\
&&&wt12&\phantom{0}0.812 (P$\uparrow$K$\uparrow$D$\uparrow$L$\uparrow$M$\uparrow$)&\phantom{0}0.785 (K$\uparrow$D$\uparrow$L$\uparrow$M$\uparrow$)&\phantom{0}0.794 (K$\uparrow$D$\uparrow$L$\uparrow$M$\uparrow$)&\phantom{0}0.786 (K$\uparrow$D$\uparrow$L$\uparrow$M$\uparrow$)&\phantom{0}0.819 (P$\uparrow$K$\uparrow$D$\uparrow$L$\uparrow$M$\uparrow$)&\phantom{0}0.786 (K$\uparrow$D$\uparrow$L$\uparrow$M$\uparrow$)\\
&&&wt13&\phantom{0}0.749 (K$\uparrow$D$\uparrow$L$\uparrow$M$\uparrow$)&\phantom{0}0.745 (K$\uparrow$D$\uparrow$L$\uparrow$M$\uparrow$)&\phantom{0}0.755 (K$\uparrow$D$\uparrow$L$\uparrow$M$\uparrow$)&\phantom{0}0.736 (K$\uparrow$D$\uparrow$L$\uparrow$M$\uparrow$)&\phantom{0}0.766 (P$\uparrow$K$\uparrow$D$\uparrow$L$\uparrow$M$\uparrow$)&\phantom{0}0.751 (K$\uparrow$D$\uparrow$L$\uparrow$M$\uparrow$)\\
&&&wt14&\phantom{0}0.773 (K$\uparrow$D$\uparrow$L$\uparrow$M$\uparrow$)&\phantom{0}0.767 (K$\uparrow$D$\uparrow$L$\uparrow$M$\uparrow$)&\phantom{0}0.766 (K$\uparrow$D$\uparrow$L$\uparrow$M$\uparrow$)&\phantom{0}0.768 (K$\uparrow$D$\uparrow$L$\uparrow$M$\uparrow$)&\phantom{0}0.785 (P$\uparrow$K$\uparrow$D$\uparrow$L$\uparrow$M$\uparrow$)&\phantom{0}0.777 (K$\uparrow$D$\uparrow$L$\uparrow$M$\uparrow$)\\
\midrule
\multirow{6}{*}{\textit{HRel}-\textit{Rel}}&
\multirow{6}{*}{\phantom{0}8.4\%}&
\multirow{6}{*}{257}
&wt09&\phantom{0}0.535 (D$\downarrow$l$\uparrow$)&\phantom{0}0.539 (L$\uparrow$)&\phantom{0}0.543 (L$\uparrow$)&\phantom{0}0.541 (L$\uparrow$)&\phantom{0}0.539 (L$\uparrow$)&\phantom{0}0.546 (p$\uparrow$L$\uparrow$)\\
&&&wt10&\phantom{0}0.585 (D$\uparrow$L$\uparrow$)&\phantom{0}0.575 (D$\uparrow$L$\uparrow$)&\phantom{0}0.580 (D$\uparrow$L$\uparrow$)&\phantom{0}0.568 (p$\downarrow$d$\uparrow$L$\uparrow$M$\downarrow$)&\phantom{0}0.581 (D$\uparrow$L$\uparrow$)&\phantom{0}0.571 (d$\uparrow$L$\uparrow$)\\
&&&wt11&\phantom{0}0.533 (D$\downarrow$M$\downarrow$)&\phantom{0}0.565 (P$\uparrow$K$\uparrow$D$\downarrow$L$\uparrow$M$\downarrow$)&\phantom{0}0.570 (P$\uparrow$K$\uparrow$D$\downarrow$L$\uparrow$M$\downarrow$)&\phantom{0}0.548 (k$\uparrow$D$\downarrow$L$\uparrow$M$\downarrow$)&\phantom{0}0.552 (P$\uparrow$K$\uparrow$D$\downarrow$L$\uparrow$M$\downarrow$)&\phantom{0}0.584 (P$\uparrow$K$\uparrow$D$\downarrow$L$\uparrow$)\\
&&&wt12&\phantom{0}0.680 (P$\uparrow$K$\uparrow$D$\uparrow$L$\uparrow$M$\uparrow$)&\phantom{0}0.637 (K$\uparrow$D$\uparrow$L$\uparrow$M$\uparrow$)&\phantom{0}0.658 (P$\uparrow$K$\uparrow$D$\uparrow$L$\uparrow$M$\uparrow$)&\phantom{0}0.651 (K$\uparrow$D$\uparrow$L$\uparrow$M$\uparrow$)&\phantom{0}0.675 (P$\uparrow$K$\uparrow$D$\uparrow$L$\uparrow$M$\uparrow$)&\phantom{0}0.641 (K$\uparrow$D$\uparrow$L$\uparrow$M$\uparrow$)\\
&&&wt13&\phantom{0}0.553&\phantom{0}0.581 (K$\uparrow$d$\uparrow$L$\uparrow$M$\uparrow$)&\phantom{0}0.580 (K$\uparrow$D$\uparrow$L$\uparrow$M$\uparrow$)&\phantom{0}0.557&\phantom{0}0.572 (k$\uparrow$M$\uparrow$)&\phantom{0}0.587 (K$\uparrow$D$\uparrow$L$\uparrow$M$\uparrow$)\\
&&&wt14&\phantom{0}0.598 (p$\uparrow$K$\uparrow$D$\uparrow$L$\uparrow$M$\uparrow$)&\phantom{0}0.589 (K$\uparrow$D$\uparrow$L$\uparrow$M$\uparrow$)&\phantom{0}0.570 (K$\uparrow$)&\phantom{0}0.599 (K$\uparrow$D$\uparrow$L$\uparrow$M$\uparrow$)&\phantom{0}0.596 (P$\uparrow$K$\uparrow$D$\uparrow$L$\uparrow$M$\uparrow$)&\phantom{0}0.584 (K$\uparrow$D$\uparrow$L$\uparrow$m$\uparrow$)\\
\midrule
\multirow{6}{*}{\textit{Rel}-\textit{NRel}}&
\multirow{6}{*}{63.5\%}&
\multirow{6}{*}{290}
&wt09&\phantom{0}0.672 (p$\uparrow$K$\uparrow$D$\uparrow$L$\uparrow$M$\uparrow$)&\phantom{0}0.665 (K$\uparrow$D$\uparrow$L$\uparrow$M$\uparrow$)&\phantom{0}0.667 (K$\uparrow$D$\uparrow$L$\uparrow$M$\uparrow$)&\phantom{0}0.670 (p$\uparrow$K$\uparrow$D$\uparrow$L$\uparrow$M$\uparrow$)&\phantom{0}0.679 (P$\uparrow$K$\uparrow$D$\uparrow$L$\uparrow$M$\uparrow$)&\phantom{0}0.671 (P$\uparrow$K$\uparrow$D$\uparrow$L$\uparrow$M$\uparrow$)\\
&&&wt10&\phantom{0}0.795 (K$\uparrow$D$\uparrow$L$\uparrow$M$\uparrow$)&\phantom{0}0.802 (K$\uparrow$D$\uparrow$L$\uparrow$M$\uparrow$)&\phantom{0}0.806 (P$\uparrow$K$\uparrow$D$\uparrow$L$\uparrow$M$\uparrow$)&\phantom{0}0.805 (p$\uparrow$K$\uparrow$D$\uparrow$L$\uparrow$M$\uparrow$)&\phantom{0}0.807 (P$\uparrow$K$\uparrow$D$\uparrow$L$\uparrow$M$\uparrow$)&\phantom{0}0.807 (P$\uparrow$K$\uparrow$D$\uparrow$L$\uparrow$M$\uparrow$)\\
&&&wt11&\phantom{0}0.778 (K$\uparrow$D$\uparrow$L$\uparrow$M$\uparrow$)&\phantom{0}0.779 (K$\uparrow$D$\uparrow$L$\uparrow$M$\uparrow$)&\phantom{0}0.788 (P$\uparrow$K$\uparrow$D$\uparrow$L$\uparrow$M$\uparrow$)&\phantom{0}0.775 (K$\uparrow$D$\uparrow$L$\uparrow$M$\uparrow$)&\phantom{0}0.792 (P$\uparrow$K$\uparrow$D$\uparrow$L$\uparrow$M$\uparrow$)&\phantom{0}0.787 (P$\uparrow$K$\uparrow$D$\uparrow$L$\uparrow$M$\uparrow$)\\
&&&wt12&\phantom{0}0.728 (K$\uparrow$D$\uparrow$L$\uparrow$M$\uparrow$)&\phantom{0}0.730 (K$\uparrow$D$\uparrow$L$\uparrow$M$\uparrow$)&\phantom{0}0.724 (K$\uparrow$D$\uparrow$L$\uparrow$M$\uparrow$)&\phantom{0}0.726 (K$\uparrow$D$\uparrow$L$\uparrow$M$\uparrow$)&\phantom{0}0.737 (P$\uparrow$K$\uparrow$D$\uparrow$L$\uparrow$M$\uparrow$)&\phantom{0}0.729 (K$\uparrow$D$\uparrow$L$\uparrow$M$\uparrow$)\\
&&&wt13&\phantom{0}0.705 (P$\uparrow$K$\uparrow$D$\uparrow$L$\uparrow$M$\uparrow$)&\phantom{0}0.685 (K$\uparrow$D$\uparrow$L$\uparrow$M$\uparrow$)&\phantom{0}0.696 (P$\uparrow$K$\uparrow$D$\uparrow$L$\uparrow$M$\uparrow$)&\phantom{0}0.693 (K$\uparrow$D$\uparrow$L$\uparrow$M$\uparrow$)&\phantom{0}0.714 (P$\uparrow$K$\uparrow$D$\uparrow$L$\uparrow$M$\uparrow$)&\phantom{0}0.697 (P$\uparrow$K$\uparrow$D$\uparrow$L$\uparrow$M$\uparrow$)\\
&&&wt14&\phantom{0}0.707 (P$\downarrow$K$\uparrow$D$\uparrow$L$\uparrow$M$\uparrow$)&\phantom{0}0.705 (P$\downarrow$K$\uparrow$D$\uparrow$L$\uparrow$M$\uparrow$)&\phantom{0}0.717 (K$\uparrow$D$\uparrow$L$\uparrow$M$\uparrow$)&\phantom{0}0.698 (P$\downarrow$K$\uparrow$D$\uparrow$L$\uparrow$M$\uparrow$)&\phantom{0}0.715 (K$\uparrow$D$\uparrow$L$\uparrow$M$\uparrow$)&\phantom{0}0.717 (K$\uparrow$D$\uparrow$L$\uparrow$M$\uparrow$)\\
\bottomrule
\end{tabular}
}
\end{table*}


\label{sec.discussion}
\begin{figure}[ht!]
  \centering
    \includegraphics[width=0.26\textwidth]{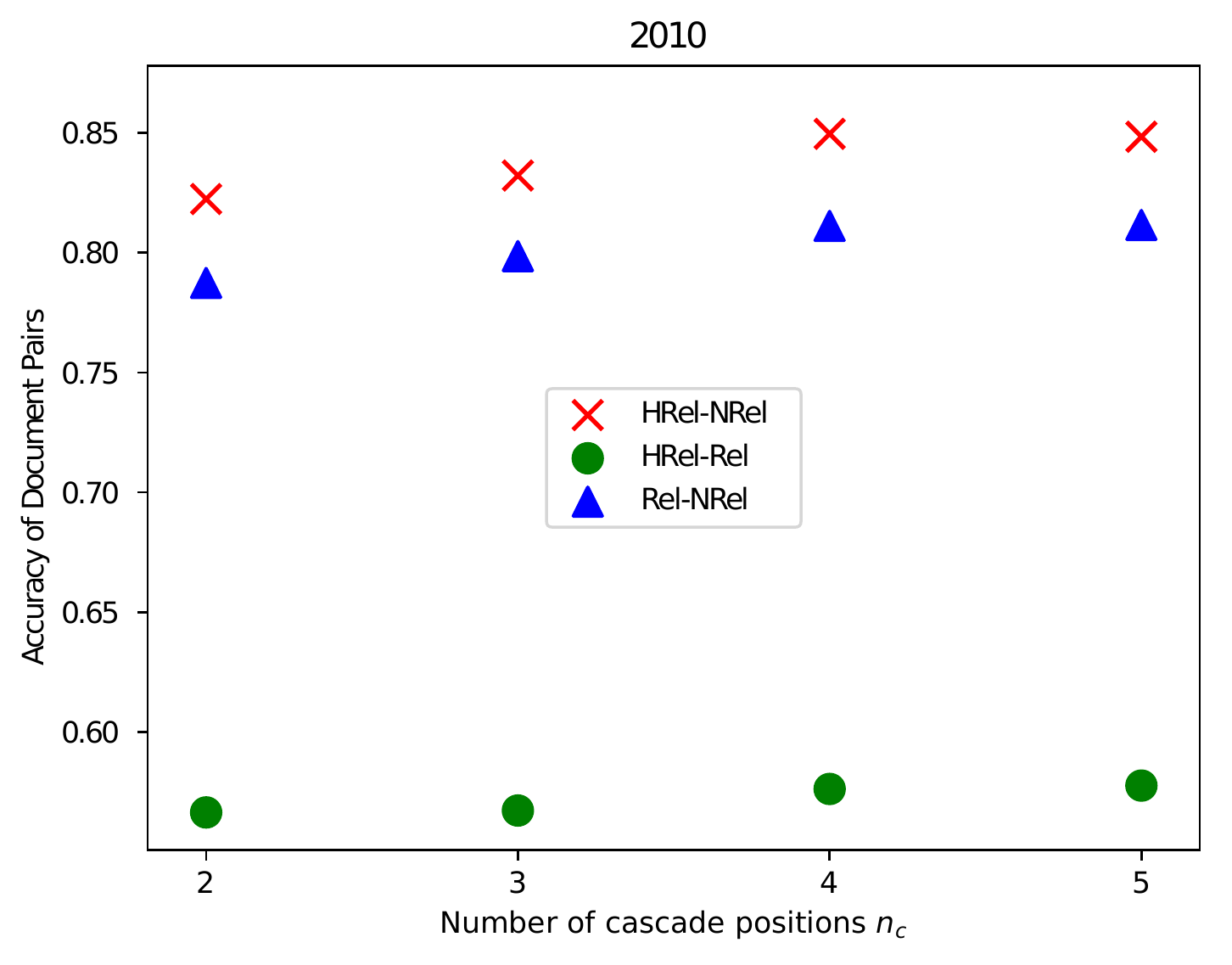}
  \caption{The accuracy on document pairs when using different number of cascade positions $n_c$
  for the cascade k-max pooling layer.}
  \label{fig.cascade}
\end{figure}

\begin{figure}[ht!]
  \centering
    \includegraphics[width=0.26\textwidth]{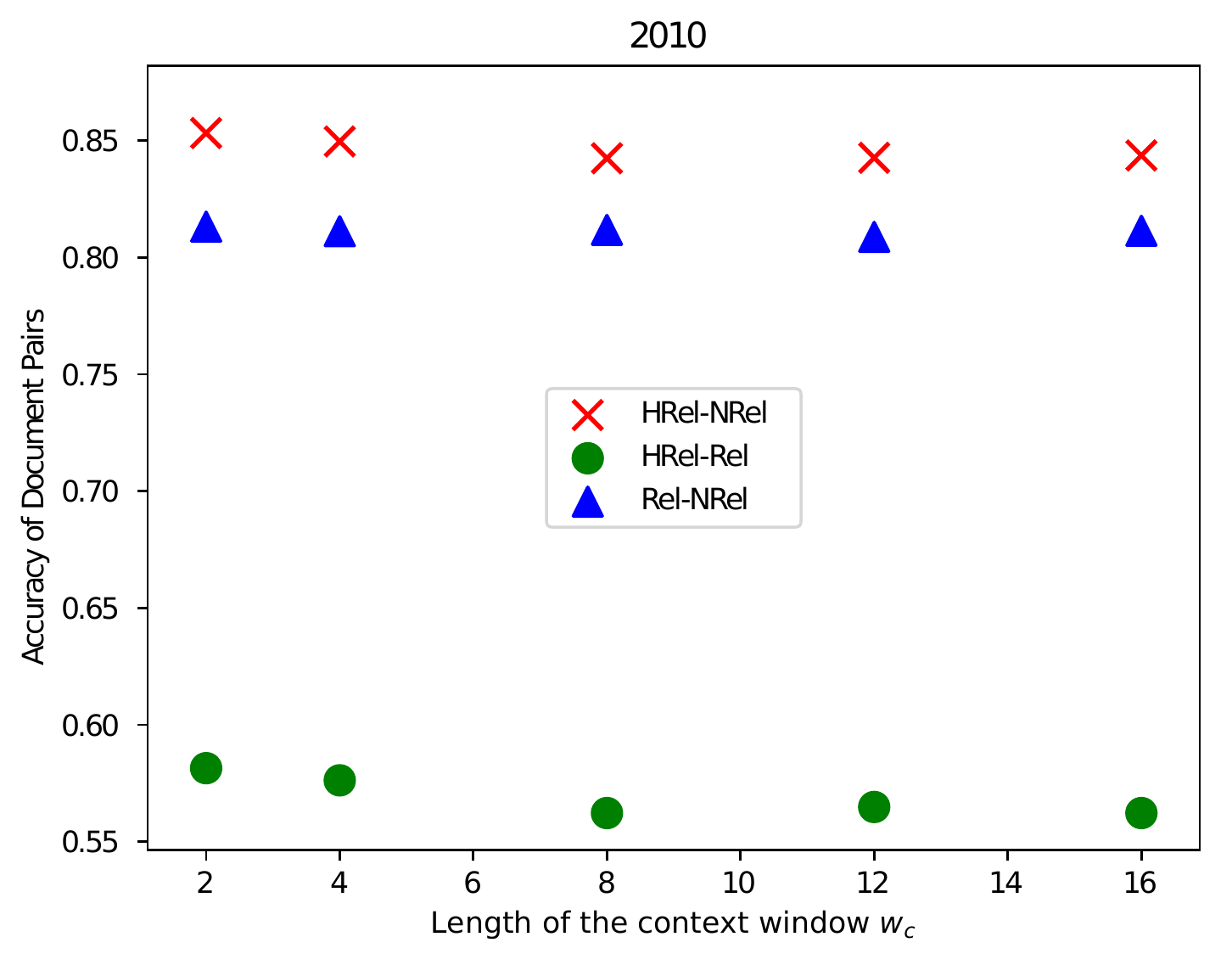}
    \caption{The accuracy on document pairs  when varying the size of the context window $w_c$
    for the disambiguation component.}
  \label{fig.context}
\end{figure}

\section{Discussion} 
\subsection{Ablation Analysis}
\label{sec.analysis}
In this section, we attempt to gain further insights about the usefulness of 
the proposed model components, 
namely, the cascade k-max pooling (C), the disambiguation (D) and the shuffling combination (S) layer,
by drawing comparisons among different model variants. As 
mentioned, the \text{PairAccuracy} benchmark is the most comprehensive 
due to its inclusion of all document pairs 
and its removal of the effects of an initial ranker,
making the analysis based solely on the proposed neural models.
Therefore,
our analysis in this section mainly considers \text{PairAccuracy}.

\textbf{Effects of the individual building blocks.}
We first incorporate the proposed components
into PACRR one at time, leading to the C-PACRR, D-PACRR, and S-PACRR model variants,
which we use to examine the effects of these building blocks separately.
Table~\ref{tab.neup.docpair}
demonstrates that 
the shuffling combination (S-PACRR) alone can boost the performance on 
three different label pairs, significantly outperforming
PACRR on two to three years out of six years for all three label combinations,
and performing at least as well as PACRR on the remaining years. 
As mentioned in Section~\ref{sec.introduction},
the shuffling combination performs regularization by
preventing the model from learning query-dependent patterns.
On the other hand,
adding the C-PACRR or D-PACRR component to PACRR
actually hurts the performance on 2014 over the Rel-NRel label pair,
and only occasionally improves PACRR on other years.
Intuitively, both  building blocks introduce extra weights into PACRR,
increasing the number of nodes for combination
by adding the context vectors or 
by using multiple pooling layers,
making the model more prone to overfitting.
Such changes might be an issue when only limited training data is available.

\textbf{Joint effects of different components.}
To resolve the extra complexity introduced by the 
cascade pooling layers and the disambiguation building blocks,
we further combine these two with the shuffling component,
leading to CS-PACRR and DS-PACRR.
Meanwhile, we also investigate the joint effects between them
by examining CD-PACRR.
From Table~\ref{tab.neup.docpair},
compared with the PACRR model, 
both CS-PACRR and DS-PACRR achieve better results not only relative to C-PACRR and D-PACRR,
but also to S-PACRR. This is especially true for CS-PACRR,
which significantly outperforms PACRR on all years for HRel-NRel pairs,
and on five years for Rel-NRel pairs.
This demonstrates that both the cascade pooling and 
the disambiguation components can help only after 
introducing extra regularization to offset 
the extra complexity being introduced.
As for CD-PACRR, not surprisingly, 
it performs on a par with C-PACRR and D-PACRR, and worse than
the CS-PACRR and DS-PACRR.
Finally, we put all components together and end up with the Co-PACRR model
discussed in Section~\ref{sec.evaluation},
which performs better than C-PACRR and D-PACRR, and similar to S-PACRR, but 
occasionally worse than CS-PACRR on 2012--14.
We argue that this is due to the joint usage of the cascade k-max pooling and
the disambiguation, making the model much more complex and thereby expensive to train
like CD-PACRR, therefore requiring more training data to work well.
We note that DS-PACRR  performs better than the S-PACRR variant,
supporting our argument that the full model's decreased performance is caused
by the added complexity, and not by adding the disambiguation component itself,
and this also applies to the cascade k-max pooling layer. 
In short, we conclude that all three components can lead to improved results.
Moreover, 
we suggest that,
when limited training data is available, 
either CS-PACRR or DS-PACRR could be employed in place of Co-PACRR,
since they are less data-hungry compared with Co-PACRR.

\subsection{Tuning of Hyper-parameters}
\label{sec.tuning}
Finally, we further investigate the effects of the two hyper-parameters introduced by our proposed components,
namely, the number of cascade positions $n_c$ and the size of the context window $w_c$, which
govern the cascade k-max pooling component and the disambiguation component, respectively.
Figures~\ref{fig.cascade} and~\ref{fig.context}
show the effects of applying different $n_c$ and $w_c$ on 2010,
where the x-axis represents the configurations of the hyper-parameter,
and the y-axis represents the corresponding accuracy on document pairs.
In the case of cascade k-max pooling, we uniformly divide $[0\%,100\%]$ into $n_c$
parts, e.g., with $n_c=5$ we have $cpos=[20\%, 40\%, 60\%, 80\%, and 100\%]$.
Owing to space constraints, 
we omit the plots for other years.
From Figures~\ref{fig.cascade} and~\ref{fig.context},
we observe that the model is robust against different choices of $n_c$ and $w_c$ within the 
investigated ranges, and the trend of the accuracy relative to different choices of 
hyper-parameters is consistent among the three kinds of label pairs.
Furthermore, increasing the number of cascade positions slightly increases the accuracy,
whereas increasing the context window size past $w_c=4$ reduces the accuracy.



\section{Conclusion} 
\label{sec.conclusion}
In this work we proposed the novel Co-PACRR neural IR model that
incorporates the local and global context of matching signals into the PACRR model through the use of a
 disambiguation building block, a cascade k-max pooling layer, and 
 a shuffling combination layer.
 Extensive experiments on \textsc{Trec} Web Track data
  demonstrated the superior performance of the proposed 
  Co-PACRR model.
  Notably, the model is trained using \textsc{Trec} data consisting of about 100k training instances,
  illustrating that models performing ad-hoc retrieval can greatly benefit from architectural improvements
  as well as an increase in training data.
   As for future work, one potential direction is the combination of 
 handcrafted learning-to-rank features with the interactions learned
 by Co-PACRR, where an effective way to learn
 such features (e.g., PageRank scores) inside the neural model appears non-trivial.


\bibliographystyle{ACM-Reference-Format}
\bibliography{kai}  
\balance

\end{document}